\documentclass[10pt,journal,doublespace]{IEEEtran}

\IEEEoverridecommandlockouts                              

\usepackage{amsmath} 

\usepackage{graphicx}
\usepackage[space]{grffile}

\title{\LARGE \bf
Feedback Solution to Optimal Switching Problems with Switching Cost
}

\author{Ali Heydari$^1$%
\thanks{$^{1}$Assistant Professor of Mechanical Engineering, South Dakota School of Mines and Technology, Rapid City, SD, email: ali.heydari@sdsmt.edu.

}}

\newtheorem{Thm}{Theorem} 
\newtheorem{Rem}{Remark} 

\begin{document}

\maketitle

\pagenumbering{arabic}
\pagestyle{plain}
\thispagestyle{plain}

\begin{abstract}
The problem of optimal switching between nonlinear autonomous subsystems is investigated in this study where the objective is not only bringing the states to close to the desired point, but also adjusting the switching pattern, in the sense of penalizing switching occurrences and assigning different preferences to utilization of different modes. The mode sequence is unspecified and a switching cost term is used in the cost function for penalizing each switching. It is shown that once a switching cost is incorporated, the optimal cost-to-go function depends on the already active subsystem, i.e., the subsystem which was engaged in the previous time step. Afterwards, an approximate dynamic programming based method is developed which provides an approximation of the optimal solution to the problem in a \textit{feedback} form and for \textit{different initial conditions}. Finally, the performance of the method is analyzed through numerical examples.
\end{abstract}

\section{Introduction}
Many real-world control problems can be classified as switching problems in the sense that the system subject to control is comprised of several different modes (sometimes called subsystems) and at each instant only one of the modes can be active. A basic example of such a system is a plant equipped with on-off actuators \cite{Heydari_GNC2013}. 
The solution to such problems includes a \textit{switching schedule} which determines the number of switching, the switching instants, and the order of the active subsystems.

The developments in the field of optimal switching can be divided into different categories, two of which are nonlinear programming based methods and discretization based methods. Nonlinear programming based methods utilize the gradient of the cost with respect to the switching instants to calculate local optimal switching times using nonlinear programming \cite{Antsaklis1}-\cite{Zhao}. In these methods, the sequence of active subsystems, known as the \textit{mode sequence}, is typically selected a priori. The problem is then simplified to determining the switching instants between the modes. Discretization based methods, however, discretize the state and input space to end up with a finite number of choices \cite{Rungger, CDC_SwitchingCost}. Among the intelligent approaches to the problem, genetic algorithm and neural networks were used in Refs. \cite{Sakly} and \cite{Long}, respectively, to determine the optimal switching for one set of initial conditions. 

All the cited methods work only with a specific initial condition; each time the initial condition is changed, a new set of computations needs to be performed to find the new optimal switching instants. In order to extend the validity of the results for different initial conditions within a pre-selected set, in \cite{Axelsson} a solution was found as the local optimum in the sense that it minimizes the worst possible cost for all trajectories starting in the selected initial states set. Another drawback of majority of the methods, especially the nonlinear programming based methods, is the fact that they lead to an open loop solution.

On the other hand, approximate dynamic programming (ADP) has shown great potentials in solving conventional optimal control problems with infinite-horizon cost functions \cite{Werbos}-\cite{Vrabie_R} and also with finite-horizon cost functions \cite{Heydari_NN}-\cite{DerongLiu}. The backbone of the ADP based methods is using Bellman equation \cite{Kirk} and approximating the mapping between the states of the system and the optimal control.
These potentials motivated the author of this study to investigate the application of ADP to \textit{switching} problems in his PhD research. This was done through developing solutions to problems with \textit{fixed mode sequence} and \textit{fixed number of switching} \cite{Heydari_TNN_Switching}, \cite{Heydari_AMC}, problems with \textit{free mode sequence} and \textit{controlled subsystems} \cite{Heydari_Neurocomputing_Switching}, and also problems with \textit{free mode sequence} and \textit{autonomous subsystems} \cite{Heydari_Franklin}.
The interesting feature of these developments is the fact that they provide approximate optimal solution for a vast domain of initial conditions. Another advantage of these methods is their feedback nature. These developments, however, do not provide the designer with the ability of influencing, e.g., decreasing the number of switching. For example, in the extreme case, the solutions proposed in \cite{Heydari_Neurocomputing_Switching} and \cite{Heydari_Franklin} can lead to one switching at every single sampling time. Such a switching frequency is impracticable in many applications. Different tricks, however, are proposed in \cite{Heydari_Neurocomputing_Switching} and \cite{Heydari_Franklin} for manipulating the switching frequency. These remedies lead to deviation of the solution from optimality and can potentially destabilize the system. Moreover, these developments assume a cost function which is independent of the active mode, hence, the designer cannot assign different costs (or preferences) to different modes. 

In an independent study in utilizing ADP for solving optimal switching problems, the authors of \cite{Zhang} proposed a method for solving switching problems with finite-horizon cost functions.
The proposed method, however, inherits the curse of dimensionality from dynamic programming, in the sense that, at each iteration of the learning process as many cost-to-go functions as the number of subsystems raised to the power of the iteration number should be learned. For example, for a three mode system, at 100th iteration, the number of functions subject to learning is $3^{100}$. Moreover, the proposed training algorithm is based on a selected initial state.

Another investigation for solving switching problems using ADP was recently reported in \cite{Ferrari_Switching} with a different approach. However, initial conditions are assumed to be known a priori and the result does not admit penalizing each switching. These two points differentiate the work from this study.

An idea for influencing (decreasing) the number of switching is incorporating a \textit{switching cost} term in the cost function, for the purpose of penalizing each switching between the modes, \cite{CDC_SwitchingCost}. Moreover, utilizing a cost function with mode dependent terms, i.e., having a \textit{switching cost function}, leads to the desired feature of assigning different costs to different modes. These modifications, however, lead to a very important change in the characteristics and the nature of the solution. It is shown in this study that in case of penalizing each switching, the optimal cost-to-go becomes a function of the subsystem which was active at the \emph{previous} time step, i.e., the \emph{already} active subsystem. Consequently, the methods reported in \cite{Heydari_TNN_Switching}-\cite{Heydari_Franklin} fail to provide solutions to problems with such cost functions. 
Based on the developments in \cite{Heydari_TNN_Switching}-\cite{Heydari_Franklin}, this study is aimed at developing a new switching method which admits the switching cost term and the switching cost function. This is the main contribution of this paper and is carried out through a new switching law, a new neural network (NN) structure as the function approximator, and a new parameter/weight update algorithm. Afterwards, the continuity of the function subject to approximation is analyzed and certain changes in the selected NN form is proposed for satisfying the necessary condition for uniform approximation of the desired function. Finally, the performance of the method is analyzed numerically in different examples.

The closest study in the literature to the problem subject to this paper is \cite{CDC_SwitchingCost}. The differences are a) a maximum number of switching needs to be assumed, b) the state space needs to be discretized, and c) the solution needs to be calculated numerically in \cite{CDC_SwitchingCost}. In this study, however, the number of switching is free, to be obtained such that the cost function is minimized, the state vector can change continuously, and the (approximate) solution is calculated in a closed form.

The rest of this paper is organized as follows. The problem if formally presented in section II and the proposed solution is detailed in section III. Section IV discusses the online implementation of the proposed method and section V includes the numerical analyses and simulations. Finally, the conclusions are given in section VI.


\section{Problem Formulation} \label{ProblemFormulation}
The dynamics of the $M$ autonomous modes/subsystems of a switching system can be modeled using
\begin{equation}
x_{k+1}=f_{i}(x_k),k \in \mathcal{K},i \in \mathcal{I} \label{Dynamics}
\end{equation}
where $f_{i}:\Re^n \to \Re^n, \forall i \in \mathcal{I}:= \left\{1,2,...,M\right\}$, $\mathcal{K} := \left\{0,1,...,N-1\right\}$, and positive integer $n$ is the dimension of the state vector $x_k$. Sub-index $k$ in $x_k$ represents the discrete time index and sub-index $i$ in $f_i(.)$ represents the respective mode/subsystem. 
Denoting the active mode at instant $k$ with $i_k$, a \textit{switching schedule} identifies $i_k, \forall k \in \mathcal{K}$. Once a switching schedule is selected, the system can operate from the initial time $k=0$ to the fixed final time $k = N$. The problem is defined as finding a switching schedule that minimizes the cost function given by
\begin{equation}
J=\psi(x_{N},i_{N-1}) + \sum_{k=0}^{N-1}\big(Q(x_{k},i_k)+\kappa(i_{k-1},i_k)\big) \label{CostFunction}
\end{equation}
Cost function (\ref{CostFunction}) is composed of three type of terms. 
a) Piecewise convex function $\psi:\Re^n \times \mathcal{I} \to \Re$ penalizes the error between the desired state value and the actual state value at the final time and is dependent on the active mode or configuration with which the operation of the system finishes, i.e., $i_{N-1}$. 
b) Piecewise convex function $Q:\Re^n \times \mathcal{I} \to \Re$ assigns different costs to the state error (the difference between the actual value and the desired value for the state vector) during the horizon and is dependent on the active mode during the horizon.
c) Piecewise constant function $\kappa:\mathcal{I} \times \mathcal{I} \to \Re$ represents the switching cost. Each switching from mode $i_{k-1}$ to $i_k$ at time $k$ leads to the cost represented by $\kappa(i_{k-1},i_k)$, \cite{CDC_SwitchingCost}. Therefore, $\kappa(i,i)=0, \forall i \in \mathcal{I}$. 
For notational consistency in (\ref{CostFunction}), the already active mode before the start of the process, i.e., before $k=0$, is denoted with $i_{-1}$.

\begin{Rem} \label{Rem1}
Functions $\psi(.,i)$ and $Q(.,i)$ are assumed to be convex, $\forall i \in \mathcal{I}$. Moreover, no assumption on the signs of the outputs of $\psi(.,.), Q(.,.),$ and $\kappa(.,.)$, e.g., being positive semi-definite, is made and the theory developed in this study admits negative costs, i.e., rewards, as well.
\end{Rem}


\section{Proposed Solution}
The method proposed in this study for solving the problem is based on approximating the \textit{optimal cost-to-go}, i.e., the total cost from the current time to the final time, assuming the optimal decisions are made in selecting the modes for operating the system during the horizon. The optimal cost-to-go is sometimes called \textit{value function} by some researchers.
It is straightforward to see that the optimal cost-to-go is a function of the current state, i.e., $x_k$. Since the final time is fixed, the cost-to-go will depend on the current time as well. In other words, having the same current state, but a different \textit{time-to-go}, i.e., different $N-k$, may lead to a different cost-to-go \cite{Kirk}, \cite{Heydari_NN}. Note that, the dependency on $N-k$ is equivalent of dependency on $k$, because, $N$ is fixed and known.

An important observation for developing a solution to the problem defined in section \ref{ProblemFormulation} is the fact that the optimal cost-to-go also depends on the \textit{previous active subsystem}, i.e., the subsystem which was active at the previous time, in problems with switching costs. The previous active subsystem, $i_{k-1}$, is `already' active, hence, utilizing it at the current time step does not cause a switching cost, because $\kappa(i_{k-1},i_{k-1})=0$. To see this dependency one may consider the difference between the following two example scenarios in controlling a switching system: a) the previous active subsystem is the same as the subsystem that the controller wants to activate at the current time, and b) having the same time $k$ and current state $x_k$ as in case 'a', the previous active subsystem is different than the one the controller wants to activate at the current time. Comparing these two scenarios it is seen that the optimal cost-to-go will be different due to the required switching in scenario 'b' and the respective incurred switching cost. Hence, the solution and the optimal cost-to-go at each instant are dependent on the previous active subsystem, as well as on the current time and state. Considering these dependencies, one may denoted the optimal cost-to-go with $V^*_k(x_k,i_{k-1})$. Note that the sub-index $k$ in $V^*_k:\Re^n \times \mathcal{I} \to \Re$ corresponds to the time dependency of the optimal cost-to-go.

Considering this concept, it is seen that the initial condition on the active mode, that is the active mode/configuration right before the start of the operation, plays a role in the selection of $i_0$, when a switching cost is incorporated. In other words, depending on what the already active mode/configuration before the start of the process is, i.e., $i_{-1}$, the system may select a different $i_0$. Therefore, as expected, the summation included in cost function (\ref{CostFunction}) contains $i_{-1}$.

\begin{Rem}
Another way of looking at the dependency of the cost-to-go on the already active subsystem, is considering the active subsystem/mode as a \textit{state} of the system. In this case, a new state vector $X_k := [x_k^T, i_{k-1}]^T $ may be defined to represent the \textit{overall} state of the system. This approach is also compatible with the physical way of looking at the modes as different \textit{configurations} of the system. The active configuration, e.g., the position of the gear stick in a manual transmission car, is a physical state of the system.
\end{Rem}

\begin{Rem} \label{Rem_AlreadyActiveMode}
The `already' active mode should be differentiated from the `current' active mode at time $k$. The former is the mode which was utilized at the `previous' time step and is denoted with $i_{k-1}$, but, the latter is the mode which is going to be selected at the 'current' time step to operate the system from $k$ to $k+1$ and is denoted with $i_{k}$. Following this terminology, the cost-to-go depends on the `previous' (or the already active) subsystem, not on the `current' subsystem.
\end{Rem}

\subsection{Theory}
The selected cost function, Eq. (\ref{CostFunction}), leads to 
\begin{equation}
V^*_N(x_N,i_{N-1})=\psi(x_N,i_{N-1}), \forall i_{N-1} \in \mathcal{I}, \label{HJB1}
\end{equation}
and
\begin{equation}
	\begin{split}
		V^*_k(x_k,i_{k-1}) &=\psi(x_N^*,i_{N-1}^*)+ \\
		& \big(Q(x_k,i_k^*) + \kappa(i_{k-1},i_k^*)\big)+ \\
		\sum_{j=k+1}^{N-1}\big(Q(x_j^*,i_j^*) &+ \kappa(i_{j-1}^*,i_j^*)\big), 
		 \forall k \in \mathcal{K}, \forall i_{k-1} \in \mathcal{I}. \label{CostToGoFormula}
	\end{split}
\end{equation}
where the \textit{optimal} active mode at each instant $j$ is denoted with $i_j^*, \forall j \in \mathcal{K},$ and the resulting optimal future states, calculated from (\ref{Dynamics}), are denoted with $x_j^*, \forall j \in \mathcal{K}\cup\left\{N\right\}$.
Eq. (\ref{CostToGoFormula}) can be formed as a recursive equation as
\begin{equation}
	\begin{split}
		V^*_k(x_k,i_{k-1}) =Q(x_k,i_k^*) + \kappa(i_{k-1}, i_{k}^*) + \\
		V^*_{k+1}(x_{k+1}^*,i_{k}^*),\forall k \in \mathcal{K}, \forall i_{k-1} \in \mathcal{I}, \label{HJB2}
	\end{split}
\end{equation}
where $x_{k+1}^* = f_{i_k^*}(x_k)$. 
By the Bellman principle of optimality \cite{Kirk}, one has
\begin{equation}
	\begin{split}
		V^*_k(x_k,i_{k-1})&= min_{i \in \mathcal{I}} \Big( Q(x_k,i) + \kappa(i_{k-1}, i) + \\
		&V^*_{k+1}(f_i(x_k),i) \Big),\forall k \in \mathcal{K}, \forall i_{k-1} \in \mathcal{I}. \label{HJB3}
	\end{split}
\end{equation}
Moreover, the optimal mode $i_k^*$, which is also a function of $k$, $x_k$, and $i_{k-1}$, is given by 
\begin{equation}
		\begin{split}
			i_k^*(x_k,i_{k-1})=argmin_{i \in \mathcal{I}} \Big( Q(x_k,i) + \kappa(i_{k-1}, i) + \\
			V^*_{k+1}(f_i(x_k),i) \Big),\forall k \in \mathcal{K}, \forall i_{k-1} \in \mathcal{I}.
		\end{split}
\label{Optimal_i_k}
\end{equation}
In other words, $i_k^*$ is selected considering the following concerns:
\begin{itemize}
\item	Selecting $i_k^*$ leads to incurring the running cost of $Q(x_k,i_k^*)$.
\item	Selecting $i_k^*$ leads to the next state vector being $f_{i_k^*}(x_k)$.
\item	Selecting $i_k^*$ leads to the fact that at the next step, the already active subsystem will be $i_k^*$.
\item	Selecting $i_k^*$ may lead to some switching cost due to $i_k^*$ not being the same as the already active subsystem, which is $i_{k-1}$.
\end{itemize}
The first concern in addressed through the inclusion of $Q(x_k,i)$ in the minimization of Eq. (\ref{Optimal_i_k}). The second and third concerns are addressed through minimizing $V^*_{k+1}(f_i(x_k),i)$ in the right hand side of Eq. (\ref{Optimal_i_k}) with respect to its both $i$s. The fourth concern, however, is addressed through the term $\kappa(i_{k-1},i)$ subject to minimization in (\ref{Optimal_i_k}). The existence of this term in Eq. (\ref{Optimal_i_k}) confirms the fact that the selection of each mode depends on the active mode at the previous step, as expected. 

The key to the solution of the problem is the fact that if the optimal cost-to-go function $V^*_k(.,.)$ is calculated in a closed form for all $k \in \mathcal{K}$ then one can find the optimal $i_k^*$ in a \textit{feedback} form in online operation, as seen in (\ref{Optimal_i_k}). 
Motivated by the development in the ADP literature for conventional \cite{Werbos}-\cite{DerongLiu} and switching \cite{Heydari_TNN_Switching}-\cite{Heydari_Franklin} problems, it is proposed to use a neural network (NN) as a function approximator for approximating the optimal cost-to-go function. Selecting a linear-in-parameter NN, the function is approximated within a compact set $\Omega \subset \Re^n$ (representing the domain of interest) using
\begin{equation}
	\begin{split}
			W_k^T & \phi(x_k,i_{k-1}) \approx V_k^* (x_k,i_{k-1}), \\
			& \forall k \in \mathcal{K}\cup\{N\}, \forall i_{k-1} \in \mathcal{I}, \forall x_k \in \Omega, \label{NN_Structure1}
	\end{split}
\end{equation}
where the selected smooth basis functions are given by $\phi:\Re^n \times \mathcal{I} \to \Re^m$, with $m$ being a positive integer denoting the number of neurons. Unknown matrix $W_k \in \Re^m$, to be found using learning algorithms, is the \textit{weight} matrix of the network at time step $k$. Note that the time-dependency of the optimal cost-to-go function is incorporated using a NN with time-dependent weights. Moreover, the inputs to the basis functions correspond to the other dependencies of the function subject to approximation. 

Before proceeding to the training algorithm, there is a concern with the selected NN structure (\ref{NN_Structure1}) that needs to be resolved. It should be noted that NNs with continuous neurons are suitable for approximation of continuous functions \cite{Hornik_NN_Continuity}, \cite{Weierstrass_Theorem}. Otherwise, the approximation is not guaranteed to be uniform. Looking at (\ref{HJB2}), function $V_k^* (x_k,i_{k-1})$ may not be a continuous function versus $i_{k-1}$. As a matter of fact, since $i_{k-1}$ belongs to a set of discrete integers, i.e., $\mathcal{I}$, it will not change continuously, therefore, the cost-to-go function also does not continuously change versus $i_{k-1}$, unless the system is comprised of only one mode. Hence, the selected network structure given in (\ref{NN_Structure1}), with continuous basis functions $\phi(.,.)$ is not desired and a new structure should be used for implementation of the proposed solution. To remedy the problem, an innovative idea is proposed here, that is, using NNs with $i_{k-1}$ dependent weights for incorporation of $i_{k-1}$-dependency of the function subject to approximation. Let a new NN structure given by
\begin{equation}
		\begin{split}
			{W_k^{i_{k-1}}}^T & \phi(x_k) \approx V_k^*(x_k,i_{k-1}), \\
			& \forall k \in \mathcal{K}\cup\{N\}, \forall i_{k-1} \in \mathcal{I}, \forall x_k \in \Omega, \label{NN_Structure2}
		\end{split}
\end{equation}
be used, where $\phi: \Re^n \to \Re^m$ is the selected set of basis functions and $W_k^i \in \Re^m, \forall k \in \mathcal{K}\cup\{N\}, \forall i \in \mathcal{I}$, is the unknown weight matrix. Using the form given by (\ref{NN_Structure2}), the number of weights required to be trained will be multiplied by the number of subsystems, as compared to the case of using the NN form given by (\ref{NN_Structure1}). 

Following this idea, it needs to be proved that each $V_k^* (x_k,i_{k-1})$ for every given $i_{k-1}$ and $k$ is a continuous function of $x_k$. Since, each one of such functions is being approximated using a different set of weights, denoted with $W_k^{i_{k-1}}$, it can be looked at as if each function $V_k^*(.,i_{k-1})$ for every given $k$ and $i_{k-1}$ is being approximated using a separate NN denoted with ${W_k^{i_{k-1}}}^T \phi(.)$. Therefore, the proof of continuity versus $x_k$ suffices for having the desired uniform approximation capability for the NN structure given in (\ref{NN_Structure2}).
Theorem 1 proves the required continuity. While the main idea of the proof is adapted from \cite{Heydari_Franklin}, many changes are carried out to adapt the result for the cost-to-go function subject to the current study and to make the proof more rigorous.

\begin{Thm} The optimal cost-to-go or value function for the problem of minimizing cost function (\ref{CostFunction}) with respect to dynamics (\ref{Dynamics}) is a continuous function of states in every compact set $\Omega$, if functions $f_i(.), \psi(.,i),$ and $Q(.,i), \forall i \in \mathcal{I}$ are continuous.
\end{Thm}

\textit{Proof}: Function $V_N^*(.,i)$ is continuous by Eq. (\ref{HJB1}) and the continuity of $\psi(.,i), \forall i$. Assuming continuity of $V_{k+1}^*(.,i), \forall i$, if it can be shown that function $V_k^*(.,i)$ for all $i$s will be continuous, the proof is complete, by mathematical induction.

Let the scalar function $F: \Re^n \times \mathcal{I} \times \mathcal{I} \to \Re$ be defined as
\begin{equation}
F\big(x,j,i) := Q(x,i) + \kappa\big(j,i\big) + V_{k+1}^*\big(f_{i}(x), i\big), \label{Theorem1_eq1}
\end{equation}
and the piecewise constant function $i_k^*: \Re^n \times \mathcal{I} \to \mathcal{I}$ be given by
\begin{equation}
i_k^*(x,j)=argmin_{i \in \mathcal{I}} \Big( F\big(x,j,i) \Big) \label{Theorem1_eq2}.
\end{equation}
It can be seen that function $F\big(x,j,i_k^*(x,j)\big)$ is identical to $V_k^*(x,j)$ considering (\ref{HJB2}), (\ref{Theorem1_eq1}), and (\ref{Theorem1_eq2}). Therefore, continuity of $F\big(.,j,i_k^*(.,j)\big)$ for all $j$s completes the proof.

Let $\bar{x}$ be any selected point in $\Omega$, for any given $j \in \mathcal{I}$ set
\begin{equation}
\bar{i}=i_k^*(\bar{x},j). \label{The1_eq1}
\end{equation}
Select an open set $\alpha \subset \Omega$ such that $\bar{x}$ belongs to the boundary of $\alpha$ and limit
\begin{equation}
\hat{i} = \lim_{\|x-\bar{x} \| \to  0, x \in \alpha} i_k^*(x,j), \label{The1_eq2}
\end{equation}
exists, where $\|.\|$ denotes a vector norm. If $\bar{i} = \hat{i}$, for every such $\alpha$, then there exists some open set $\beta \subset \Omega$ containing $\bar{x}$ such that $i_k^*(x,j)$ is constant for all $x \in \beta$, because $i_k^*(x,j)$ only assumes integer values. In this case the continuity of $F\big(x,j,i_k^*(x,j)\big)$ at $x = \bar{x}$ follows from the fact that $F\big(x,j,i\big)$ is continuous at $x = \bar{x}$, for every fixed $i \in \mathcal{I}$, by composition. The reason is $Q(.,i)$, $f_{i}(.)$, and $V_{k+1}^*(.,i)$ are continuous functions and $\kappa\big(j, i\big)$ is a constant. Finally, the continuity of the function subject to investigation at every $\bar{x} \in \Omega$, leads to the continuity of the function in $\Omega$.

Now assume $\bar{i} \neq \hat{i}$, for some $\alpha$. From the continuity of 
$F(x,j,\hat{i})$
for the given $\hat{i}$, one has
\begin{equation}
F(\bar{x},j,\hat{i}) = \lim_{\delta x \to 0} \big(F(\bar{x} + \delta x,j,\hat{i})\big)
\label{The1_eq4}
\end{equation}
If it can be shown that, for every selected $\alpha$, one has
\begin{equation}
F(\bar{x},j,\bar{i}) = F(\bar{x},j,\hat{i}),
\label{The1_eq5}
\end{equation}
then the continuity of $F(x,j,i_k^*(x,j))$ versus $x$ follows, because from (\ref{The1_eq4}) and (\ref{The1_eq5}) one has
\begin{equation}
F(\bar{x},j,\bar{i}) = \lim_{\delta x \to 0} \big(F(\bar{x} + \delta x,j,\hat{i})\big),
\label{The1_eq6}
\end{equation}
and (\ref{The1_eq6}) leads to the continuity by definition \cite{Trench}. 
The proof that (\ref{The1_eq5}) holds is done by contradiction. Assume that for some $\bar{x}$ and some $\alpha$ one has
\begin{equation}
F(\bar{x},j,\bar{i}) < F(\bar{x},j,\hat{i}),
\label{The1_eq7}
\end{equation}
then, due to the continuity of both sides of (\ref{The1_eq7}) at $\bar{x}$ for the fixed $\bar{i}$ and $\hat{i}$, there exists an open set $\gamma$ containing $\bar{x}$, such that 
\begin{equation}
F(x,j,\bar{i}) < F(x,j,\hat{i}),\forall x \in \gamma. \label{The1_eq8}
\end{equation}
Inequality (\ref{The1_eq8}) implies that at points which are \textit{close enough} to $\bar{x}$, one has $i_k^*(x,j) \neq \hat{i}$. But, this contradicts Eq. (\ref{The1_eq2}) which implies that there always exists a point $x$ \textit{arbitrarily close} to $\bar{x}$ at which $i_k^*(x,j) = \hat{i}$. Therefore, equality (\ref{The1_eq8}) cannot hold. Now, assume that
\begin{equation}
F(\bar{x},j,\bar{i}) > F(\bar{x},j,\hat{i}),. \label{The1_eq9}
\end{equation}
Inequality (\ref{The1_eq9}) leads to $i_k^*(\bar{x},j) \neq \bar{i}$. But, this is against (\ref{The1_eq1}), hence, (\ref{The1_eq9}) also cannot hold. Therefore, (\ref{The1_eq5}) holds and hence, $F(x,j,i_k^*(x,j))$ is continuous at every $x \in \Omega$ for every fixed $j \in \mathcal{I}$. This completes the proof.  
\vspace{3 mm}

\subsection{Training Algorithms}
Selecting the NN structure, the next step is developing an algorithm for finding the unknown weights. Using Eqs. (\ref{HJB1}) and (\ref{HJB3}) the training algorithm can be derived in a \textit{backward} fashion. Considering (\ref{HJB1}), unknown $W_N$ can be obtained, for example using least squares method, as shown in \cite{Heydari_NN}. Once $W_N$ is found, Eq. (\ref{HJB3}) can be used for calculating $W_{N-1}$. Repeating this process, all the weights can be found from $k=N$ to $k=0$. The training can be done either in a \textit{batch} form or in a \textit{sequential} form. Algorithm 1 details the batch training and Algorithm 2 presents the training in the sequential form.

\setlength{\parindent}{.0cm} 
\setlength{\hangindent}{1.0cm} 
\textbf{Algorithm 1 - Batch Training}

\setlength{\parindent}{.0cm} 
\setlength{\hangindent}{1.0cm} 
Step 1: Randomly select $p$ different state vectors $x^{[q]} \in \Omega, q \in \left\{1,2,..,p\right\}$, for $p$ being a large positive integer, where $\Omega \subset \Re^n$ represents the domain of interest.

Step 2: For $j = 1$ to $M$ repeat Step 3.

\setlength{\parindent}{1cm} 
\setlength{\hangindent}{2.0cm} 
Step 3: Find $W_N^j$ such that
\setlength{\hangindent}{0cm} 
\begin{equation}
{W_N^j}^T \phi(x^{[q]}) \approx \psi(x^{[q]},j), \forall q \in \left\{1,2,..,p\right\}. \label{Alg1_Eq1}
\end{equation}

\setlength{\parindent}{0cm} 
\setlength{\hangindent}{1.0cm} 
Step 4: Set $k=N-1$.

Step 5: For $j = 1$ to $M$ repeat Step 6.

\setlength{\parindent}{1cm} 
\setlength{\hangindent}{2.0cm} 
Step 6: Find $W_k^j$ such that 
\setlength{\hangindent}{0cm} 
\begin{equation}
	\begin{split}
		{W_k^j}^T \phi(x^{[q]} ) \approx min_{i \in \mathcal{I}}  \Big( Q(x^{[q]},i) + \kappa(j, i) + \\
		{W_{k+1}^i}^T \phi\big(f_i(x^{[q]})\big)\Big), \forall q\in \left\{1,2,..,p\right\}. \label{Alg1_Eq2}
	\end{split}
\end{equation}
	
\setlength{\parindent}{.0cm} 	
\setlength{\hangindent}{1.0cm} 
Step 7: Set $k=k-1$. Go back to Step 5 until $k=0$.

\vspace{02 mm}

\setlength{\parindent}{0cm} 
\setlength{\hangindent}{0cm} 
\textbf{Algorithm 2 - Sequential Training}

\setlength{\parindent}{0cm} 
\setlength{\hangindent}{1cm} 
Step 1: For $j = 1$ to $M$ repeat Step 2.

\setlength{\parindent}{1cm} 
\setlength{\hangindent}{2cm} 
Step 2: Select an initial guess on ${W_N^j}$ and repeat Steps 3 and 4 until ${W_N^j}$ converges.

\setlength{\parindent}{1.5cm} 
\setlength{\hangindent}{2.5cm} 
Step 3: Randomly select state vector $x \in \Omega$, where $\Omega \subset \Re^n$ represents the domain of interest.

\setlength{\parindent}{1.5cm} 
\setlength{\hangindent}{2.5cm} 
Step 4: Train weight $W_N^j$ of neural network $W_N^j \phi(.)$ using input-target pair $\big\{x, \mbox{ } \psi(x,j) \big\}$.

\setlength{\parindent}{0cm} 
\setlength{\hangindent}{1cm} 
Step 5: Set $k=N-1$.

\setlength{\parindent}{0cm} 
\setlength{\hangindent}{1cm} 
Step 6: For $j = 1$ to $M$ repeat Step 7.

\setlength{\parindent}{1cm} 
\setlength{\hangindent}{2cm} 
Step 7: Select an initial guess on ${W_k^j}$ and repeat Steps 8 and 9 until ${W_k^j}$ converges.

\setlength{\parindent}{1.5cm} 
\setlength{\hangindent}{2.5cm} 
Step 8: Randomly select state vector $x \in \Omega$.

Step 9: Train $W_k^j$ using input-target pair $\Big\{x, \mbox{ } min_{i \in \mathcal{I}}  \Big( Q(x,i) + \kappa(j, i) + {W_{k+1}^i}^T \phi\big((f_i(x)\big)\Big) \Big\}$.

\setlength{\parindent}{0cm} 
\setlength{\hangindent}{1cm} 
Step 10: Set $k=k-1$. Go back to Step 6 until $k=0$.

\setlength{\parindent}{.3cm} 
\vspace{3 mm}

\begin{Rem} \label{Rem3} In order to have an idea of the computational load of the proposed method, one may consider the batch training form, Algorithm 1. The backward-in-time form of the algorithm resembles the solution to the conventional optimal control problem of discrete-time linear systems with quadratic cost functions, where, the Riccati (difference) equation needs to be evaluated for $N$ time steps to find and store the time-varying solution, for all $k \in \mathcal{K}$, \cite{Kirk}. Eq. (\ref{Alg1_Eq1}) resembles the final condition on the solution and Eq. (\ref{Alg1_Eq2}) resembles the Riccati difference equation which takes the solution corresponding to $(k+1)$ and provides the solution for time $k$. Due to the nonlinearity and the hybrid nature of the problem subject to this study, a function approximator needs to be utilized and several sample state vectors need to be selected at each evaluation of Eq. (\ref{Alg1_Eq2}), for example, to find the unknown parameters. However, it can be seen that the computational load of the algorithm grows \emph{linearly} as the number of time steps increases.
\end{Rem}

Finally, before concluding this section, it should be noted that the selection of \textit{linear-in-parameter} form for the NN, as done in (\ref{NN_Structure1}) and (\ref{NN_Structure2}), is not required for the theory developed in this study to be valid. One can utilize \textit{multi-layer perceptrons}, with $k$ and $i_{k-1}$ dependent weights, for improving the approximation capability of the NN. In this case, for example (\ref{NN_Structure2}) changes to 
\begin{equation}
	\begin{split}
				NN(& W_k^{i_{k-1}},x_k) \approx V_k^*(x_k,i_{k-1}),\\
						& \forall k \in \mathcal{K}\cup\{N\}, \forall i_{k-1} \in \mathcal{I}, \forall x_k \in \Omega, \label{MLP_NN}
	\end{split}
\end{equation}
where function $NN(.,.)$ denotes the NN mapping, with the first argument being the tunable weights of the NN and the second argument being its input.

\section{Online Control}

Once the NN is trained, it can be used for online control of the switching system. The process involves feeding the current state $x_k$ and the already active subsystem $i_{k-1}$ to equation
\begin{equation}
		\begin{split}
				i_k^*(x_k,i_{k-1}) = & argmin_{i \in \mathcal{I}}  \Big( Q(x_k,i) + \kappa(i_{k-1}, i) + \\
				& {W_{k+1}^i}^T \phi\big((f_i(x_k)\big)\Big), 
		\end{split}
		\label{OnlineScheduling}
\end{equation}
to find $i_k^*(x_k,i_{k-1})$. Note that, the minimization proposed in (\ref{OnlineScheduling}) is composed of comparing $M$ scalar values and selecting the $i \in \mathcal{I}$ corresponding to the least value. Hence, the online \textit{global} minimum can be easily found. 

The advantages of this method are numerous. Firstly, the method provides a \textit{feedback} solution, hence, it will be relatively robust toward uncertainties and disturbances. 
Secondly, no restriction is enforced on the order of the active subsystems or on the number of switching.
Thirdly, unlike the nonlinear programming based methods \cite{Antsaklis1}-\cite{Zhao} which give a local optimum, this method leads to an approximation of the global optimal solution. Note that this feature holds if the training of the NN, itself, is not stuck in a local minimum, which with the selection of linear-in-parameter NN and using convex methods like least squares, the condition is fulfilled.
Fourthly, an important feature of this method is providing optimal switching for any initial condition $x_0 \in \Omega$ as long as the resulting state trajectory lies in the domain on which the network is trained, i.e., $x_k \in \Omega, \forall k$. The reason is the cost-to-go approximation is valid when the state belongs to $\Omega$. 
Finally, the method provides a great deal of flexibility for implementation of different desired switching behaviors through admitting a general cost function with switching terms.

\section{NUMERICAL ANALYSES}

Two examples are selected for numerical investigation of the features of the proposed scheme. The source codes, in MATLAB, are available upon request.

\subsection{Example 1}
As the first example, a scalar problem with two modes, given below, is selected,
\begin{equation}
	\dot{x} = f_1(x(t)) := -x(t), \mbox{ } \dot{x} = f_2 (x(t)) := -x^3(t),
\end{equation}
with the horizon of $2 s$. For discretization of the continuous-time system, Euler forward integration, with a sampling time of $0.02 s$ is selected which leads to $N=100$. The selected cost function is 
\begin{equation}
		J= 5x_{100}^2 + \sum_{k=0}^{99}\kappa(i_{k-1},i_k),
\end{equation}
where
\begin{equation}
			\mbox{ }
			\kappa(i_{k-1},i_k) = \left\{ 
					\begin{array}{ll}
					 0 & \mbox{if } i_{k-1}=i_k\\
					 0.1 & \mbox{if } i_{k-1} \neq i_k \\
					\end{array}
					\right.
\end{equation}
Therefore, while the objective is bringing the state to close to zero, a cost of $0.1$ is assumed for each switching. Hence, the controller should make a tradeoff between the cost due to the error in the state at the final time, and the cost due to switching. Considering the subsystems’ dynamics, both are stable. Comparing the derivatives of the state, however, subsystem 1 has a faster convergence rate when $|x| < 1$. But, when $|x|>1$, subsystem 2 leads to a faster convergence of the state to the origin. Therefore, assuming there was no switching cost, the optimal solution would have been

\begin{equation}
				i_k^* = \left\{ 
			  \begin{array}{ll}
         1 & \mbox{if } |x| < 1\\
				 2 & \mbox{if } |x|>1 \\
				\end{array}. \right.
				 \label{Ex1_OptimalSol}
\end{equation} 
In comparing the neurocontroller results with Eq. (\ref{Ex1_OptimalSol}), it should be noted that Eq. (\ref{Ex1_OptimalSol}) is, loosely speaking, a ``pointwise'' optimal active mode, in the sense that it does not account for the cost of switching.

\begin{figure}[t!]
	\centering
		\includegraphics[width=0.9\columnwidth]{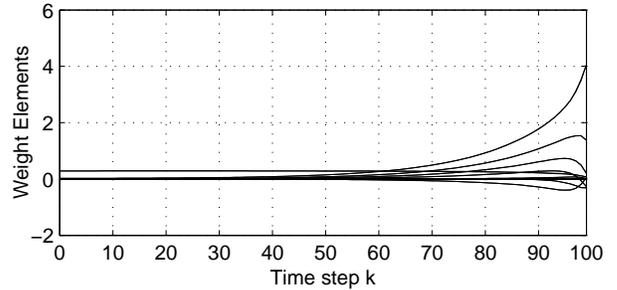}	
		\caption{History of elements of the weight of the trained NN.}
	\label{Fig0}
\end{figure}

The basis functions were selected as polynomials $x^j$, where $j \in \{1,2,...,14\}$. The accuracy of the approximation capability of the NN can be adjusted by the selection of the order of the polynomials. The training was done over the domain of $\Omega=[-2 \mbox{ } 2]$ in a sequential form and the weight were observed to converge in 1000 iterations. 
The resulting weight histories for the NN are plotted in Fig. \ref{Fig0}. As expected, the weights are observed to be time-dependent, which represents the time-dependency of the cost-to-go. 

\begin{figure}[t!]
	\centering
		\begin{tabular}{cc}
				\includegraphics[width=0.9\columnwidth]{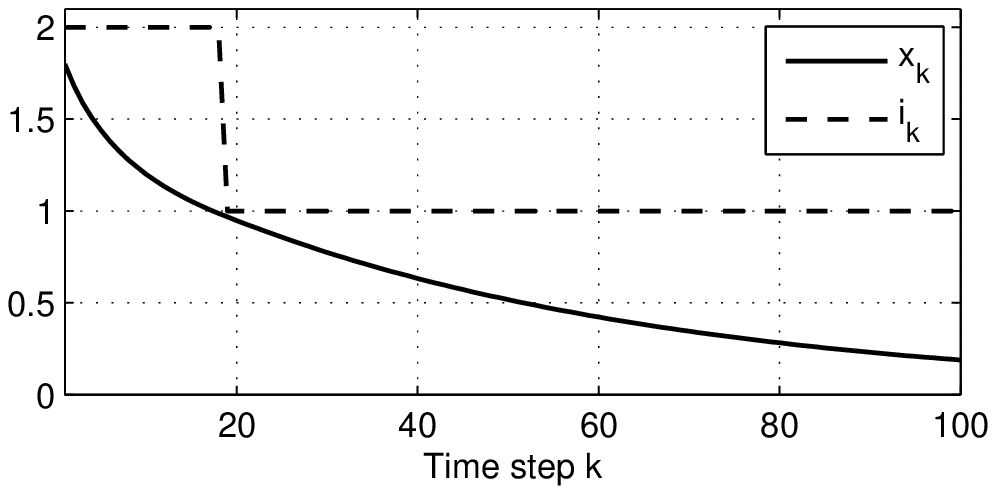} \\ \small{a: Initial mode $i_{-1}=2$.} \\
				\includegraphics[width=0.9\columnwidth]{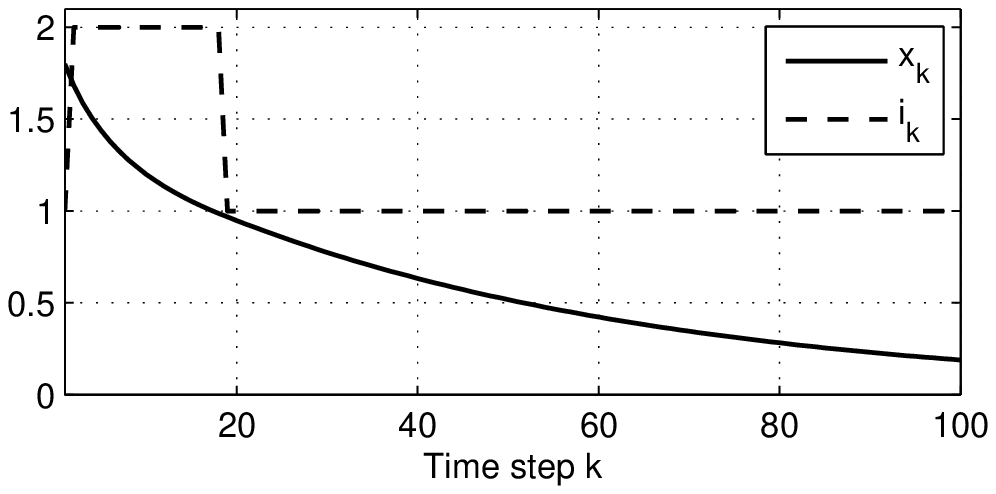} \\ \small{b: Initial mode $i_{-1}=1$.} \\
		\end{tabular}
	\caption{History of state and active mode for initial condition $x_0=1.8$.}
	\label{Fig1}
\end{figure}

After training the neurocontroller, initial condition $x_0 = 1.8$ was simulated using the developed method, for both cases of $i_{-1}=2$ and $i_{-1}=1$, i.e., the initial active mode being either subsystem 2 or subsystem 1. The results are given in Fig. \ref{Fig1}. 
Considering the case of $i_{-1}=2$, the utilized mode in the initial 18 time steps is \textit{optimal} based on Eq. (\ref{Ex1_OptimalSol}). Moreover, once the state becomes less than 1, if switching to mode 1 is eventually needed, i.e., the switching cost is unavoidable, then the switching should happen immediately based on Eq. (\ref{Ex1_OptimalSol}), as done in Fig. \ref{Fig1}.a. Comparing the cost-to-go 0.187 corresponding to Fig. \ref{Fig1}.a with the cost-to-go of staying with mode 2 without any switching, which turned out to be 1.15, it is seen that the switching was required and the controller has provided optimal solution to the problem.
Such an argument can be made for $i_{-1}=1$ as well to analyze its optimality. The cost-to-go of the schedule given in Fig. \ref{Fig1}.b turned out to be 0.197 which is less than the cost-to-go of operating mode 1 for the entire time (that is 0.297). Therefore, both switching, conducted in Fig. \ref{Fig1}.b, were needed. Comparing the switching times with Eq. (\ref{Ex1_OptimalSol}), the result given in Fig. \ref{Fig1}.b is also optimal.

\begin{figure}[t!]
	\centering
		\begin{tabular}{cc}
				\includegraphics[width=0.9\columnwidth]{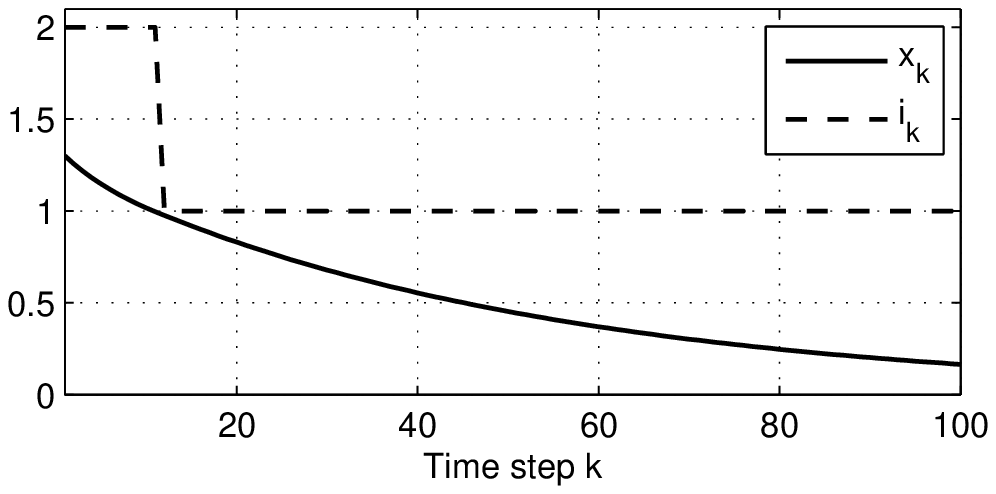} \\
				\small{a: Initial mode $i_{-1}=2$.}  \\
				\includegraphics[width=0.9\columnwidth]{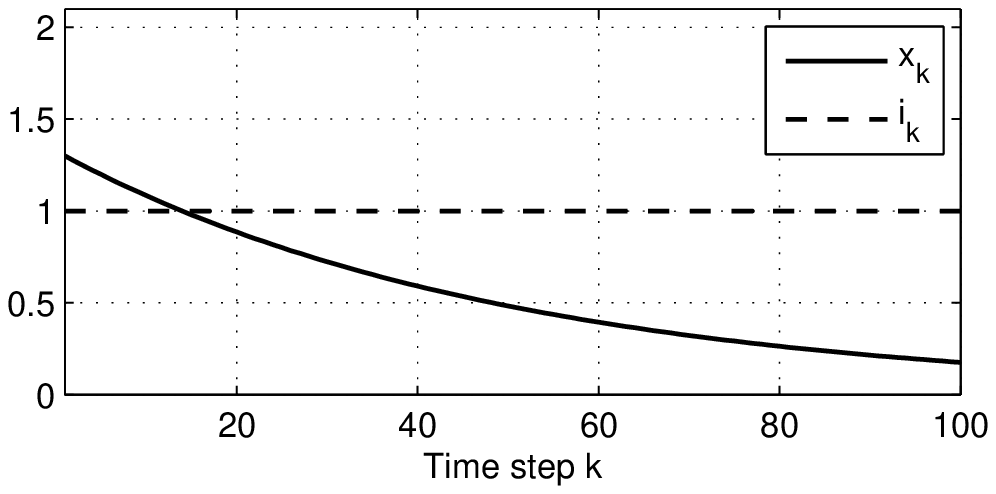} \\
				\small{b: Initial mode $i_{-1}=1$.} \\
		\end{tabular}
	\caption{History of state and active mode for initial condition $x_0=1.3$.}
	\label{Fig2}
\end{figure}

Note that the controller is able to provide optimal control for a vast number of initial conditions as long as the resulting state trajectory lies within $\Omega$. From the dynamics of the subsystems it can be seen that selecting any $x_0 \in \Omega$ leads to $x_k \in \Omega, \forall k \in \mathcal{K}$. Therefore, the trained network can optimally control any initial condition $x_0 \in \Omega$. The initial condition $x_0= 1.3$ was selected next. The results are presented in Fig. \ref{Fig2}. While Fig. \ref{Fig2}.a and comparing its cost-to-go, 0.146, with the cost-to-go of no switching at all, 1.084,
show that the controller has optimally switched between the modes per Eq. (\ref{Ex1_OptimalSol}), an interesting observation can be made from Fig. \ref{Fig2}.b. As seen in this figure, once the initial active mode is not ``pointwise'' optimal, but, the initial condition is not large enough such that switching to the pointwise optimal mode leads to an enough \textit{reward} to cover the cost of switching, the controller stays with the initial mode and skips the switching. This can be confirmed by comparing the cost-to-go of the case of switching from mode 1 to mode 2 at the very beginning and switching back to mode 1 right when $x$ becomes smaller than 1, which turned out to be 0.156, with the cost of staying with mode 1 for the entire time, that is 0.155. Since the cost-to-go of the former case is more than that of the latter case, no switching was needed and the controller has optimally controlled the new initial condition.

A very important feature of the solution is the fact that this method does not \textit{postpone} switching instants, as the remedy proposed in \cite{Heydari_Franklin} does. If a switching is eventually needed, then it should happen at the \textit{best} time without spending time with operating the non-optimal mode. For example, in Figs. \ref{Fig1}.a and \ref{Fig2}.a, where the initial active mode is 2 and a switching is eventually needed, the switching has happened right at the best time, i.e., the time that the state became less than 1.

\begin{figure}[t!]
	\centering
		\begin{tabular}{cc}
				\includegraphics[width=0.9\columnwidth]{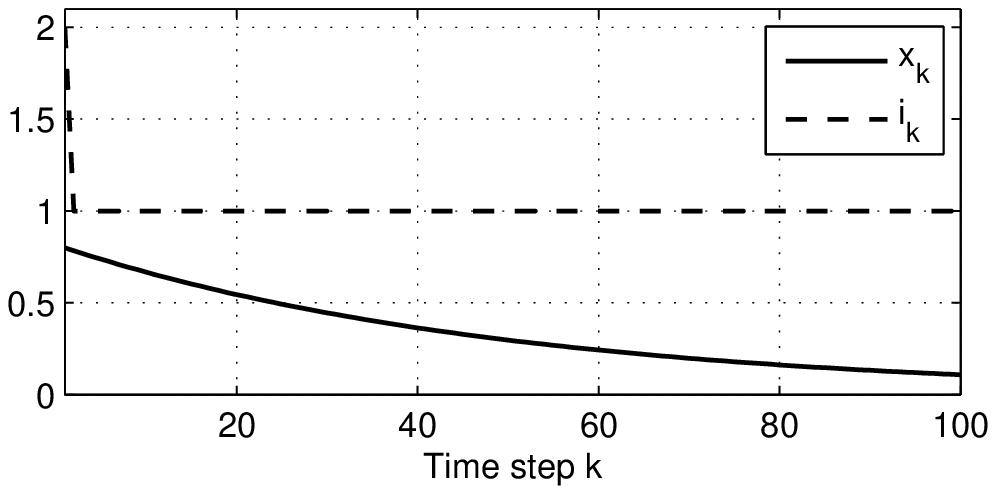} \\
				\small{a: Initial mode $i_{-1}=2$.} \\
				\includegraphics[width=0.9\columnwidth]{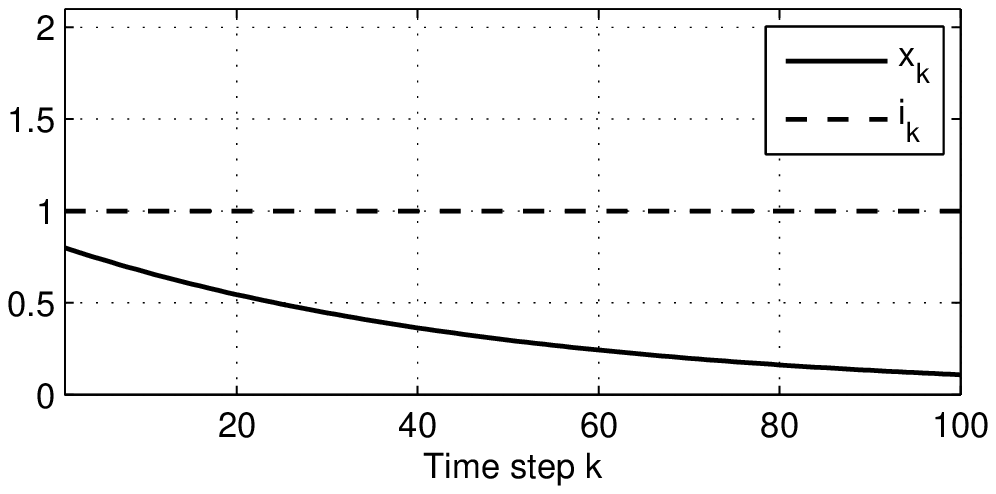} \\
				\small{b: Initial mode $i_{-1}=1$.} \\
		\end{tabular}
	\caption{History of state and active mode for initial condition $x_0=0.8$.}
	\label{Fig3}
\end{figure}

Finally, the initial condition $x_0 = 0.8$ is simulated for both the two initial active modes, and the results are depicted in Fig. \ref{Fig3}. Considering the history of the active modes, the neurocontroller has optimally controlled the system through either not switching at all, or switching immediately, considering the state histories.

\subsection{Example 2}
A nonlinear second order system with three modes, simulated in \cite{Two_Tank} and \cite{Heydari_Franklin}, is selected as the second example. The objective of this problem is controlling the fluid level in a two-tank setup. The fluid flow into the upper tank can be adjusted through a valve which has three positions: fully open, half open, and fully closed. Each tank leaks fluid with a rate proportional to the square root of the height of the fluid in the respective tank. The upper tank leaks into the lower tank, and the lower tank leaks to the outside of the setup. Representing the fluid height in the upper tank with scalar $y$ and in the lower tank with scalar $z$, the dynamics of the state vector $x=[y,z]^T$ are given by the following three modes, corresponding to the three positions of the valve,
\begin{equation}
	\begin{split}
		\dot{x}=f_1(x) := \left[ 
		\begin{array}{l}
			-\sqrt{y}\\
			\sqrt{y}-\sqrt{z}\\
		\end{array}
		\right], \\
		\mbox{ }
		\dot{x}=f_2(x) := \left[ 
		\begin{array}{l}
			-\sqrt{y}+0.5\\
			\sqrt{y}-\sqrt{z}\\
		\end{array}
		\right], \\
		\mbox{ }
		\dot{x}=f_3(x) := \left[ 
		\begin{array}{l}
			-\sqrt{y}+1\\
			\sqrt{y}-\sqrt{z}\\
		\end{array}
		\right] .
	\end{split}
\end{equation}

The objective is forcing the fluid level in the lower tank, i.e., $z$, to track constant value $0.5$. Selecting the control horizon $5 s$ the problem was discretized using sampling time of $0.025 s$, therefore $N = 200$. Then, cost function (\ref{CostFunction}) was selected for evaluating the performance of the method in both decreasing the number of switching and also for assigning certain preferences in utilization of some modes. 
The basis functions for this example were selected as polynomials $y^p z^q$, where non-negative integers $p$ and $q$ are such that $p+q \leq 8$. This selection led to 45 neurons. Domain $\Omega = \{[y,z]^T \in \Re^2: 0 \leq y < 1, 0 \leq z < 0.8 \}$ was used for the training and batch training scheme was selected, such that, at each stage 100 random states were selected based on Algorithm 1. 

\begin{figure}[ptbh]
		\centering
		\includegraphics[width=0.9\columnwidth]{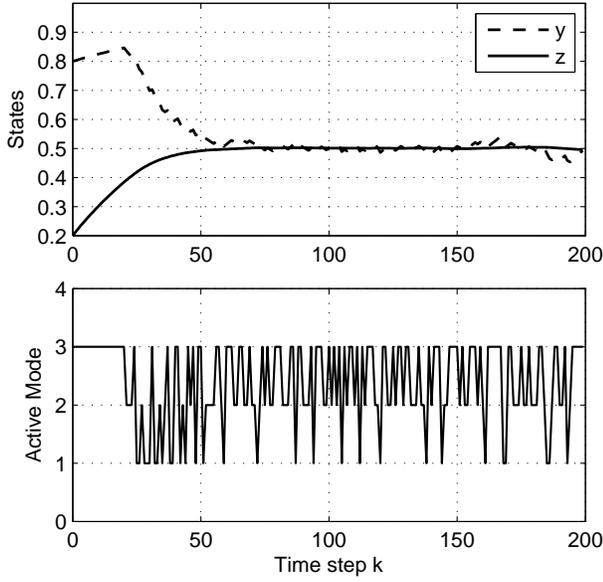} 
		\caption{Simulation result for $x_0=[0.8, 0.2]^T, \kappa_0=0,$ and cost function terms given in Eq. (\ref{Example2_CostTerms1}).} 
		\label{Ex2_Fig1}
\end{figure}

\begin{figure}[ptbh]
		\centering
		\includegraphics[width=0.9\columnwidth]{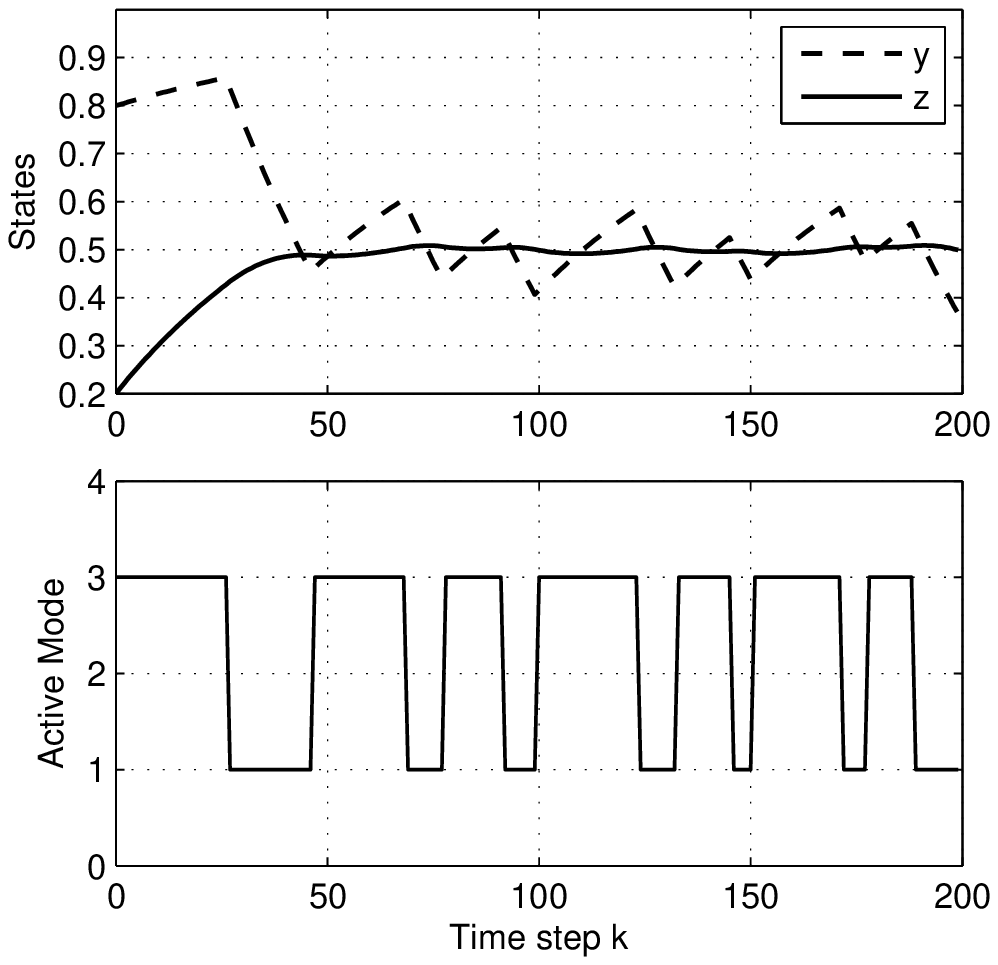} 
		\caption{Simulation result for $x_0=[0.8, 0.2]^T, \kappa_0=0.001,$ and cost function terms given in Eq. (\ref{Example2_CostTerms1}).} 
		\label{Ex2_Fig2}
\end{figure}

\begin{figure}[ptbh]
		\centering
		\includegraphics[width=0.9\columnwidth]{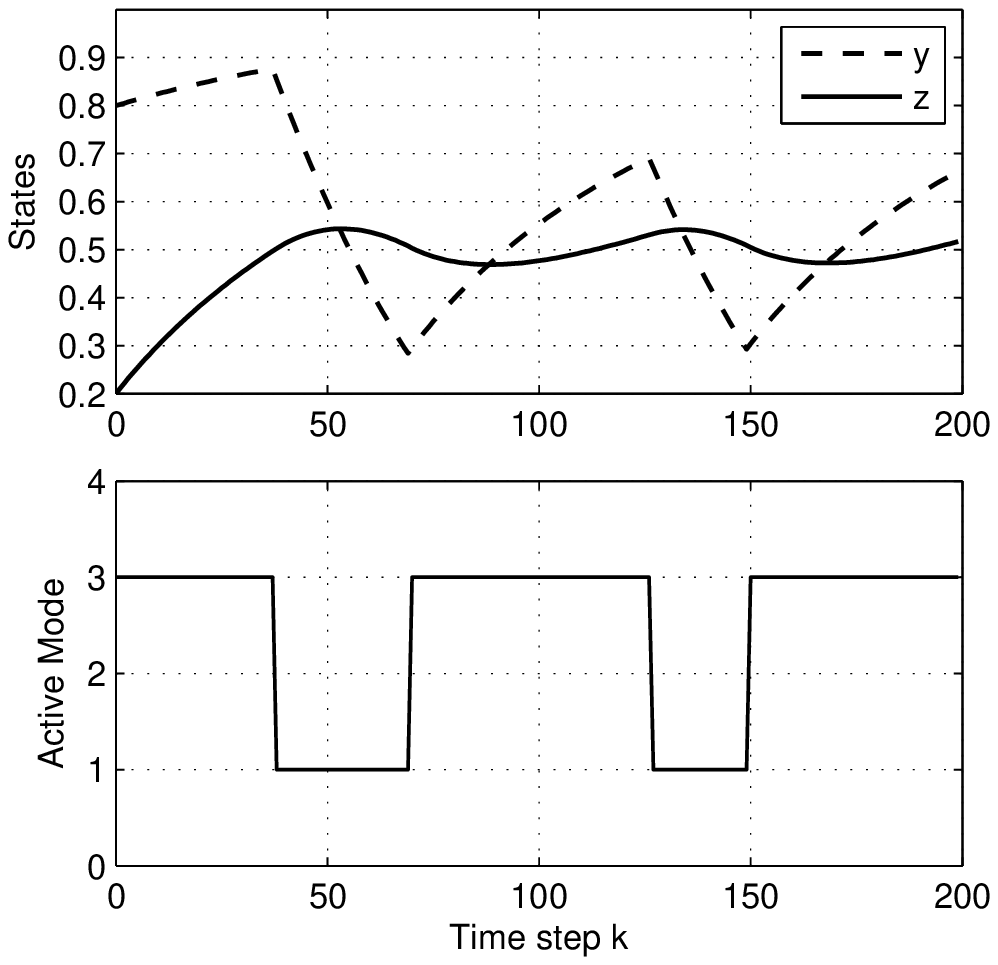} 
		\caption{Simulation result for $x_0=[0.8, 0.2]^T, \kappa_0=0.01,$ and cost function terms given in Eq. (\ref{Example2_CostTerms1}).} 
		\label{Ex2_Fig3}
\end{figure}

Initially the following values were selected for the cost function
\begin{equation}
	\begin{split}
			\psi(x_{N},i_{N-1}) = 0.25(z_N-0.5)^2, \forall i_{N-1} \in \mathcal{I}, \\
			Q(x_{k},i_k) = 0.25(z_k-0.5)^2, \forall i_{k} \in \mathcal{I}, \\
			\kappa(i_{k-1},i_k) = \left\{ 
					\begin{array}{ll}
					 0 & \mbox{if } i_{k-1}=i_k\\
					 \kappa_0 & \mbox{if } i_{k-1} \neq i_k \\
					\end{array}, \right.
		\label{Example2_CostTerms1}
	\end{split}
\end{equation}
with $\kappa_0=0$.
As seen, such a cost function does not assign any cost to switching and does not differentiate between the modes. The training was observed to take almost 16 seconds, when $N=200$, in a machine with CPU Intel Core i7, 3.4 GHz running MATLAB 2013a. Afterwards, initial condition $x_0 = [0.8,0.2]^T$, simulated in \cite{Two_Tank} and \cite{Heydari_Franklin}, was used to determine the optimal solution. Selecting $i_{-1}=3$, the results are given in Fig. \ref{Ex2_Fig1}. The method did an excellent job controlling the fluid level of the lower tank by tracking the desired value. This perfect tracking was achievable, however, through high frequency switching between the three modes. 
Next, the switching cost $\kappa_0=0.001$ was used with terms given in (\ref{Example2_CostTerms1}). The training was re-done using the new cost function and the simulation results in controlling the same initial condition are shown in Fig. \ref{Ex2_Fig2}. It is seen that the incorporated switching cost has effectively lowered the number of switching, compared with Fig. \ref{Ex2_Fig1}.
Assigning higher cost to switching, like $\kappa_0=0.01$, further decreases the number of switching while tracking 0.5, as shown in Fig. \ref{Ex2_Fig3}. 

\begin{figure}[ptbh]
		\centering
		\includegraphics[width=0.9\columnwidth]{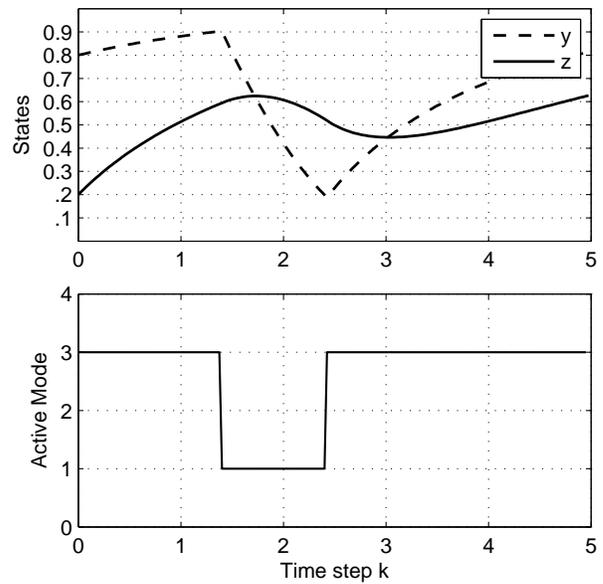} 
		\caption{Simulation result for $x_0=[0.8, 0.2]^T, \kappa_0=0,$ and cost function terms given in Eq. (\ref{Example2_CostTerms1}) with applied threshold $0.01$ proposed in \cite{Heydari_Franklin}.}
		\label{Ex2_Fig8}
\end{figure}

The application of switching costs for decreasing the number of switching resembles the idea utilized in \cite{Heydari_Franklin} and called \textit{Threshold Remedy}, in which no switching cost was incorporated, and hence, the optimal cost-to-go was approximated as a standard function of only the state and the time, as in non-switching problems \cite{Heydari_NN}. The training also was done without including a switching cost. In online control process, however, a threshold, similar to the switching cost used in this study, was applied. This was done in the sense that, the reward of switching to another mode must be higher than a certain threshold in order for the controller to switch. 
While this remedy can help in certain conditions, the performance deviates as the threshold becomes large. The reason is, the threshold is not accounted for in the learning process and hence, the result given in \cite{Heydari_Franklin} is not optimal considering the applied threshold. To see this, one may compare Fig. \ref{Ex2_Fig3} with the results given in Fig. \ref{Ex2_Fig8}, where the former is the result of the proposed method in this study with $\kappa_0=0.01$ and the latter is the result obtained from \cite{Heydari_Franklin} with the equal threshold of 0.01, in lieu of the switching cost. As seen, the tracking is much more accurate in Fig. \ref{Ex2_Fig3} versus Fig. \ref{Ex2_Fig8}. Even considering the cost due to the extra switches in Fig. \ref{Ex2_Fig3}, the cost-to-go corresponding to Fig. \ref{Ex2_Fig3} turned out to be 0.340 which is less than the cost-to-go corresponding to Fig. \ref{Ex2_Fig8}, i.e., 0.468. High threshold values can potentially lead to unreliability of the result of the method in \cite{Heydari_Franklin}. This problem does not exists with the method presented here due to the fact the the switching cost is incorporated in the derivation and the solution is optimal with respect to it.

Once the performance of the controller in applying switching costs is analyzed, the cost function terms $\psi(.,.)$ and $Q(.,.)$ are modified to assign certain \textit{mode preferences} in the operation of the system. The mode preference is using modes 2 and 3 more often \textit{during} the operation and \textit{finishing} the operation preferably with mode 1. This was done through selecting the following cost function terms 
\begin{equation}
		\begin{split}
		& \psi(x_{N},i_{N-1}) = \left\{ 
			  \begin{array}{ll}
         0.25(z_N-0.5)^2 -10 , & \mbox{if } i_{N-1}=1\\
         0.25(z_N-0.5)^2, & \mbox{if } i_{N-1} \neq 1\\
				\end{array}, \right. \\
				& Q(x_{k},i_k) = \left\{ 
			  \begin{array}{ll}
         0.25(z_k-0.5)^2 + 0.01 , & \mbox{if } i_{k}=1\\
         0.25(z_k-0.5)^2, & \mbox{if } i_{k} \neq 1\\
				\end{array}	\right.
  \label{Example2_CostTerms2}		
			\end{split}
\end{equation}
along with the switching cost $\kappa(i_{k-1},i_k)$ given in (\ref{Example2_CostTerms1}) with $\kappa_0=0$. Note that the negative cost -10 assigned to $\psi(.,1)$ provides the controller with rewards if the last active mode is 1. Also, the positive cost 0.01 assigned to $Q(.,1)$ penalizes the usage of mode 1 during the horizon. Moreover, it should be noted that the method presented in this study does not require the cost function terms to be positive semi-definite, see Remark \ref{Rem1}. Hence, one can select smooth functions with negative values as well as positive semi-definite functions.

\begin{figure}[ptbh]
		\centering
		\includegraphics[width=0.9\columnwidth]{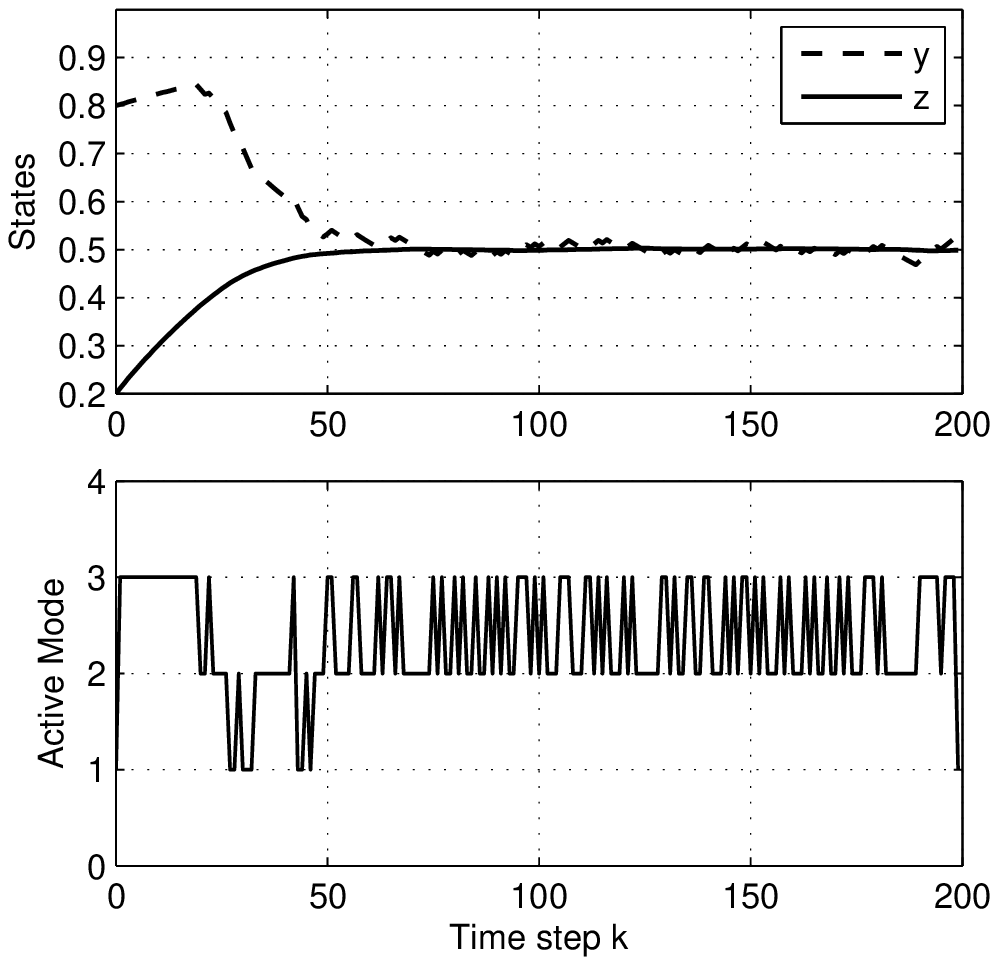} 
		\caption{Simulation result for $x_0=[0.8, 0.2]^T, \kappa_0=0,$ and cost function terms given in Eq. (\ref{Example2_CostTerms2}).} 
		\label{Ex2_Fig4}
\end{figure}

\begin{figure}[tbhp]
		\centering
		\includegraphics[width=0.9\columnwidth]{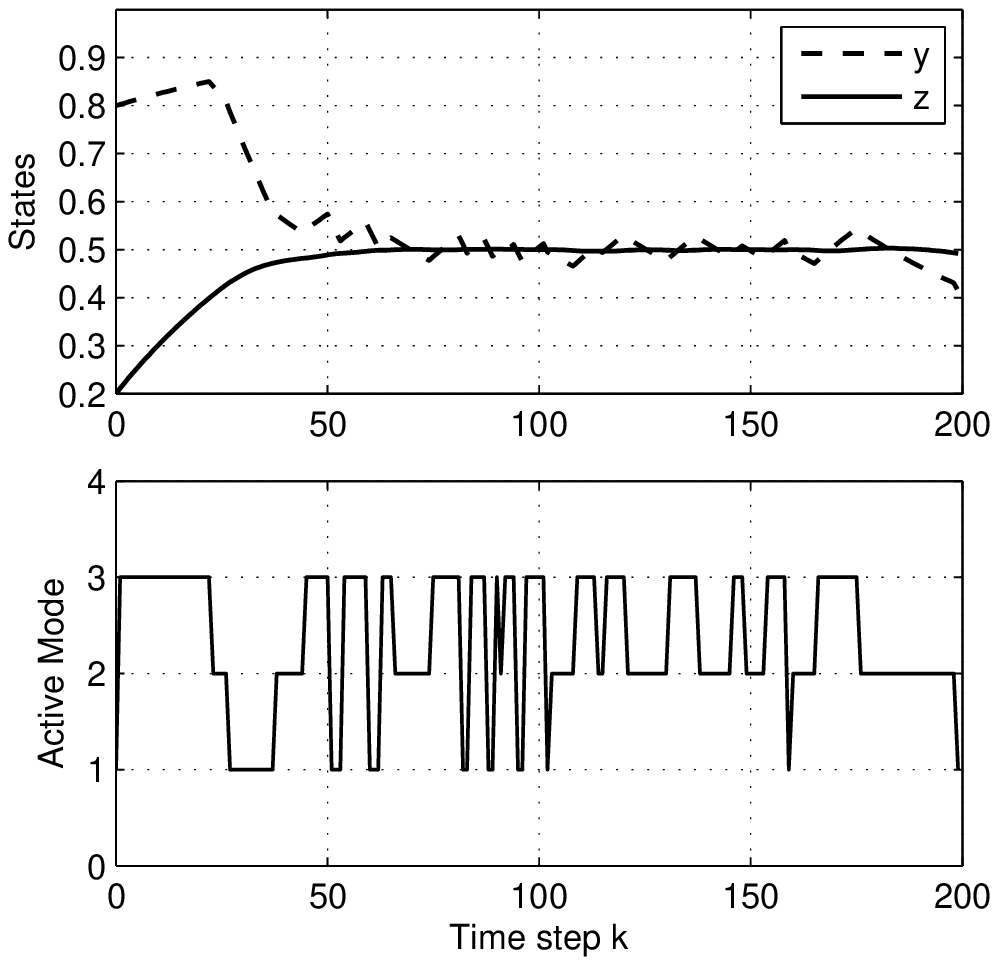} 
		\caption{Simulation result for $x_0=[0.8, 0.2]^T, \kappa_0=0.0002,$ and cost function terms given in Eq. (\ref{Example2_CostTerms2}).} 
		\label{Ex2_Fig6}
\end{figure}

Having trained the neurocontroller based on the new cost function terms, the results for controlling the same initial $x_0$, with $i_{-1}=1$, are presented in Fig. \ref{Ex2_Fig4}. Comparing this figure with Fig. \ref{Ex2_Fig1}, it is seen that the new controller has effectively decreased the number of instants that mode 1 is used. Also, looking at the final active mode in Fig. \ref{Ex2_Fig4}, the controller has been successful in finishing the operation with mode 1, as desired. As another simulation, the cost function terms given in (\ref{Example2_CostTerms2}) with the switching cost $\kappa_0=0.0002$ is used for training the network and the simulation results are given in Fig. \ref{Ex2_Fig6}. This figure demonstrates the capability of the method in both lowering the number of switching and assigning different preferences to utilization of different modes.

\begin{figure}[ptbh]
		\centering
		\includegraphics[width=0.9\columnwidth]{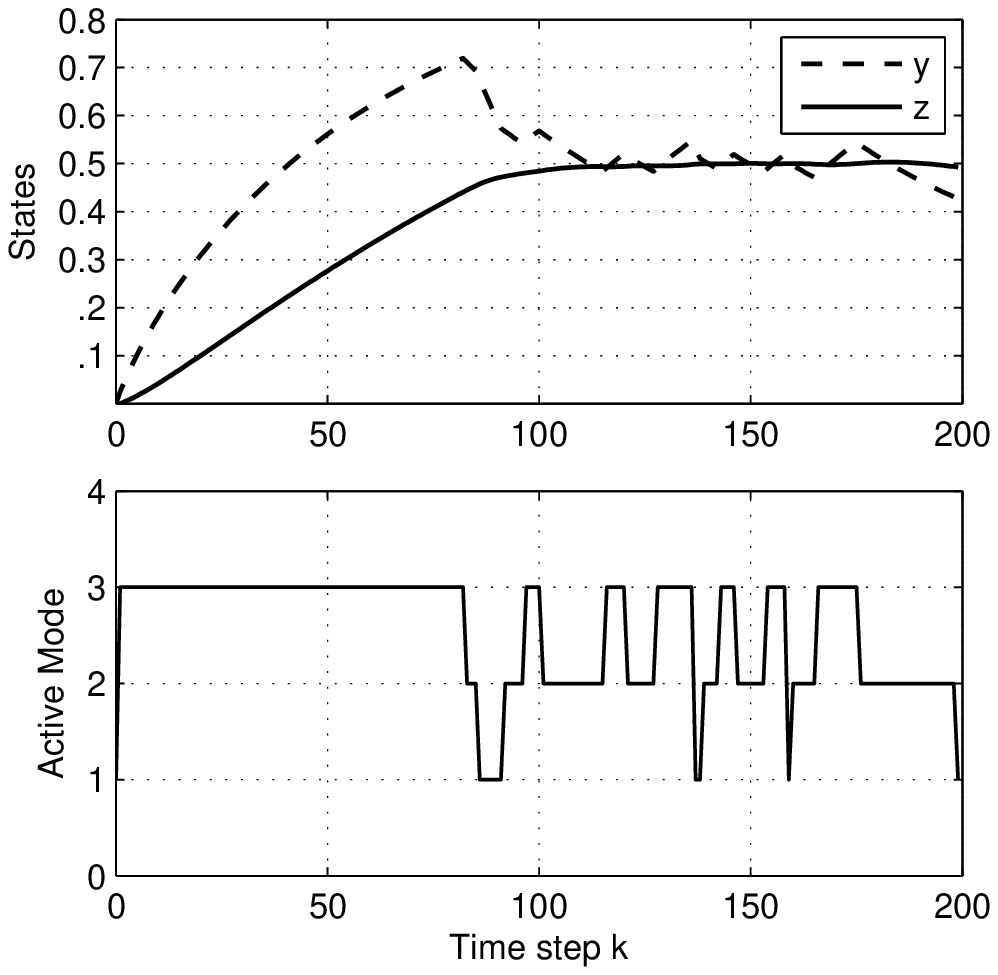} 
		\caption{Simulation result for $x_0=[0, 0]^T, \kappa_0 = 0.0002$, and cost function terms given in Eq. (\ref{Example2_CostTerms2}).} 
		\label{Ex2_Fig7}
\end{figure}

\begin{figure}[ptbh]
		\centering
		\includegraphics[width=0.9\columnwidth]{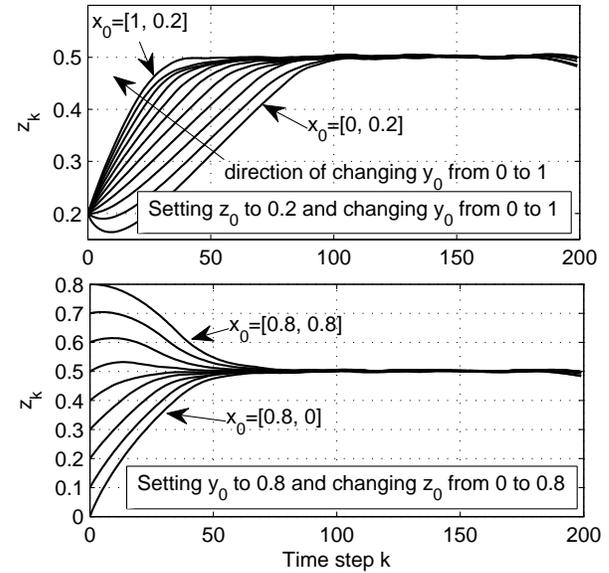} 
		\caption{Simulation result for different initial conditions with $\kappa_0 = 0.0002$, and cost function terms given in Eq. (\ref{Example2_CostTerms2}).} 
		\label{Ex2_Fig10}
\end{figure}

Next, the capability of the neurocontroller in controlling different initial conditions within $\Omega$ is investigated. 
A new initial condition, namely $x_0 = [0,0]^T$ is utilizes as the last simulation and the last trained network (without re-training) is used for controlling it. The results, given in Fig. \ref{Ex2_Fig7}, show the capability of the controller in controlling the new initial condition with the desired manner without any need for retraining. In order to furthermore evaluate this capability, several different initial conditions are selected and the resulting histories for $z_k$, which is the variable of interest, from simulating each of the initial conditions are presented in Fig. \ref{Ex2_Fig10}. In these plots, initially $z_0$ is fixed at 0.2 and $y_0$ is changed from 0 to 1 in steps of 0.1, and then, $y_0$ is fixed at 0.8 and $z_0$ is changed from 0 to 0.8, in steps of 0.1. The plots show that the controller has effectively controlled all these different initial conditions through tracking the desired constant value 0.5, without re-training.

\begin{figure}[ptbh]
		\centering
		\includegraphics[width=0.9\columnwidth]{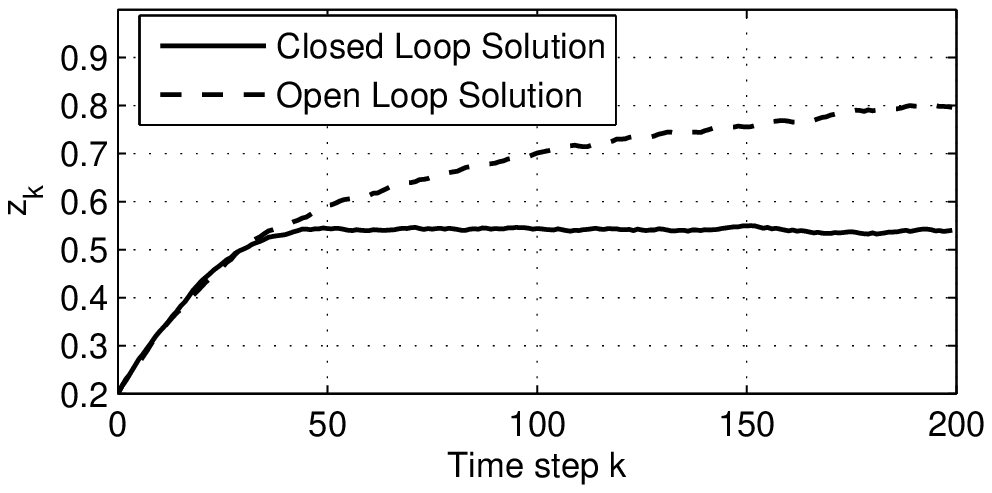} 
		\vspace{-4 cm}
		\caption{Simulation result under random time-varying disturbance for $x_0=[0.8, 0.2]^T, \kappa_0 = 0.0002$, and cost function terms given in Eq. (\ref{Example2_CostTerms2}).} 
		\label{Ex2_Fig11}
\end{figure}

Finally, the closed-loop feature of the proposed method is evaluated in terms of its moderate robustness. A time-varying random disturbance is introduced after the training phase in the online operation and is assumed to be uniformly changing between 0 and 0.005, acting as additive terms on both state elements. The resulting history for the variable of interest $z_k$, is depicted in Fig. \ref{Ex2_Fig11} through the solid plot. Moreover, the history if the system was operated using an open loop solution is also included, through the dash plot in the same figure. As seen, the system operated in the open loop fashion, that is, when the mode sequence in Fig. \ref{Ex2_Fig6}, which was calculated under no disturbance situation is applied, leads to instability of the system, but, the closed from solution handles the disturbance with a slight performance degradation in terms of a small steady state error. This demonstrates the desired feature of the proposed method.


\section{CONCLUSIONS}
The problem of finding the optimal switching schedule between different modes of a dynamical system with a cost function which admits incorporation of switching costs as well as assigning different costs to different modes was investigated and the framework of approximate dynamics programming was used with the idea of approximating the optimal cost-to-go for solving it. It was shown that for such problems the cost-to-go function is not only a function of the current state and time, but also, a function of the subsystem which was active at the previous time step. 
It was shown that the developed technique can effectively provide (approximate) optimal solutions to the problems with different initial conditions in a feedback form. The real-time computational burden of the method is as small as evaluating as many scalar-valued functions as the number of subsystems.

\end{document}